# Graphene-Silicon Schottky diodes for photodetection*

Antonio Di Bartolomeo, Giuseppe Luongo, Laura Iemmo, and Filippo Giubileo

*Abstract—* We present the optoelectronic characterization of two graphene/silicon Schottky junctions, fabricated by transferring CVD-graphene on flat and nanotip-patterned n-Si substrates, respectively. We demonstrate record photo responsivity, exceeding 2.5 A/W under white light, which we attribute to the contribution of charges photogenerated in the surrounding region of the flat junction or to the internal gain by impact ionization caused by the enhanced field on the nanotips.

## I. Introduction

The graphene/silicon (Gr/Si) junction has been the subject of an intense research activity both for the easy fabrication and for the variety of phenomena that it allows to study. It offers the opportunity to investigate new fundamental physics at the interface between a 2D semimetal and a 3D semiconductor, and holds promises for a new generation of graphene-based devices such as photodetectors, solar cells and chemical-biological sensors [1].

## II. Features of the Gr/Si junction

A Gr/Si junction with defect-free interface exhibits rectifying current-voltage (I-V) characteristics, which are the result of the formation of a Schottky barrier, as in traditional metal-semiconductor (M/S) Schottky diodes. Moreover, a Gr/Si junction presents features that are absent in M/S diodes. The vanishing density of states at the graphene Dirac point enables Fermi level tuning and hence Schottky barrier height modulation by a single anode-cathode bias [1,2]. Fig. 1 shows the energy bands alignment in a Gr/n-Si junction and the modulation of the Schottky barrier height, $\Phi_B$, which is decreased (increased) by the reverse (forward) bias; graphene is assumed p-type, such being the common doping of air-exposed samples [3].

When the Gr/Si junction is used as a photodiode, graphene acts not only as anti-reflecting and transparent conductive layer for charge transport to the external circuit, but it functions also as active material for light absorption and electron-hole generation and separation. Graphene absorbs about 2.3% of the incident light, independently of the wavelength, from near ultraviolet to near infrared [4]. Although most of the incident light is converted to photocharge into Si, the absorbance in graphene enables detection of photons with Si sub-bandgap energy through internal photoemission over the Schottky barrier [5]. Photocharges injected over the Schottky barrier (Fig. 2 (a)), under high reverse bias, can be accelerated by the electric field in the depletion region of the diode and cause avalanche multiplication by scattering with the Si lattice, thus enabling internal gain.

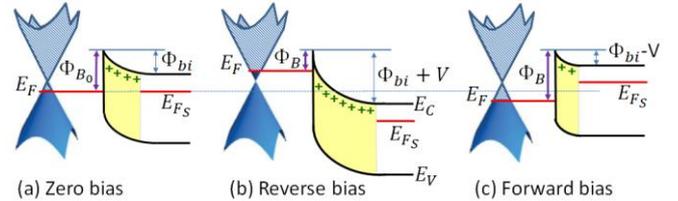

Figure 1. Gr/Si junction energy bands alignment at (a) zero, (b) reverse and (c) forward bias. Due to the low density of states of graphene, the Schottky barrier height, $\Phi_B$, is modulated by the applied bias. $\Phi_{bi}$ and $V$ are the built-in potential and the external bias of the junction; $E_F$ and $E_{F_S}$ are the Fermi energies of graphene and Si, while $E_C$ and $E_V$ are the lowest and highest levels of the Si conduction and valence band, respectively.

Successful application of the Gr/Si junction for photodetection at Si sub-bandgap energies, for instance at 1550 nm which is of interests for optical communication, has been reported, with responsivity up to 0.37A/W in reverse bias [6].

The Gr/Si junction forms the ultimate ultra-shallow junction, which is ideal to detect light absorbed very close to the Si surface, such as near- and mid-ultraviolet, which generates charges in a less than 20 nm deep layer. The location of the Gr/Si junction reduces the diffusion path and prevents the severe surface recombination that occurs in traditional pn junctions. Indeed, detection with high responsivity of 1.14 A/W has been demonstrated for the wavelength range from 200 to 400 nm [7].

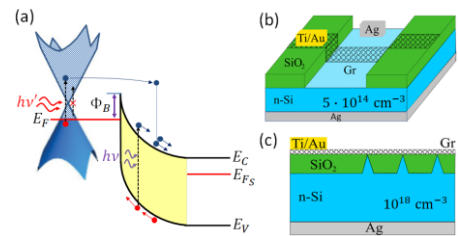

Figure 2. (a) Photodection in a Gr/Si junction. Photons with enery lower than the Si bandgap, $E_G = E_C - E_V$, but higher than the Schottky barrier $\Phi_B$ ($\Phi_B < h\nu < E_G$) can be absorbed in graphene. Emitted over the Schottky barrier, such electrons can originate avalanche multiplication through impact ionization. (b) Graphene on flat Si substrate ("flat Gr/Si" junction) and (c) graphene on patterned Si ("Gr/Si-tips" junction).

In this paper, we report the electrical characterization and the photoresponse of two types of Gr/Si devices, shown in Figs

*Research supported by POR Campania FSE 2014–2020, Asse III Ob. specifico l4, Avviso pubblico decreto dirigenziale n. 80 del 31/05/2016 and LR num. 5/2002 Finanziamento progetti annualità 2008, Prot. 2014, 0293185, 29/04/2014.

A. Di Bartolomeo is with the Physics Department, University of Salerno, and with CNR-SPIN, Fisciano 84084, Salerno, Italy (corresponding author: phone: +39089969189;fax: +39089969658; e-mail: adibartolomeo@unisa.it).

G. Luongo and L. Iemmo are with the Physics Department, University of Salerno, and with CNR-SPIN, Fisciano 84084, Salerno, Italy (e-mails: giluongo@unisa.it; liemmo@unisa.it).

Filippo Giubileo is with CNR-SPIN, 84084, Fisciano, Salerno, Italy (e-mail: filippo.giubileo@spin.cnr.it).

2 (b) and 2 (c), which are fabricated on flat and nanotips terminated Si surfaces. In the following, we will refer to them as the "flat Gr/Si" [8,9] and the "Gr/Si-tips" [10] device, respectively.

### III. DEVICE FABRICATION AND MEASUREMENTS

The fabrication involves the patterning of the substrate and the transfer of a monolayer CVD-graphene from Cu foils by a wet etch method [11]. For the flat Gr/Si device [8], the substrate preparation implies the opening of a 10 μm wide slit in the capping $SiO_2$ layer by reactive ion etching (RIE), followed by buffered HF cleaning, immediately prior to graphene transfer. The size of the transferred graphene foil is about $1 \times 0.4$ cm$^2$, implying that most of the graphene sheet extends over the 60 nm thick $SiO_2$ layer. Fig. 3 (a) shows graphene both on Si and $SiO_2$; the micro-Raman spectra of Fig. 3 (b) confirm high-quality monolayer graphene everywhere.

The fabrication of the Gr/Si-tips devices [10, 12] starts with RIE etching of the Si-tips followed by a two-step dielectric deposition (USG and BPSG films), the chemical mechanical polishing (CMP) process to uncover the tips up to the desired diameter, and the transfer of graphene (Fig. 3 (c) and 3 (d)).

The electrical measurements are performed by a Keithley 4200 SCS using graphene (contacted with a Ti/Au layer [13]) as the anode and the Si substrate as the cathode, in a Janis ST500 cryogenic probe station under controlled pressure (typically 50 mbar) and temperature. Photoconduction is investigated under white light from a LED array with adjustable intensity up to 5 mW/cm$^2$ and spectrum ranging from 420 to 720 nm.

### IV. FLAT GR/SI JUNCTION: RESULTS AND DISCUSSION

The I-V characteristic (Fig. 4 (a)) of the flat Gr/Si junction, at room temperature, shows a rectifying behavior typical of a Schottky diode that can be fitted by the equation:

$$I = I_0 \left[\exp\left(\frac{qV}{nkT}\right) - 1\right] \quad (1)$$

with

$$I_0 = AA^*T^2 \exp\left(-\frac{\Phi_{B0}}{kT}\right) \quad (2)$$

where $I_0$ is the reverse saturation current, n the ideality factor, A the area of the Gr/Si junction, $A^* = 112$ Acm$^{-2}$K$^{-2}$ is the effective Richardson constant for n-Si, T is the temperature and $\Phi_{B0}$ is the Schottky barrier height at zero bias. A more realistic model (Fig. 4 (c)) includes a series resistance, $R_s$, cumulative of all the resistive paths in the device and the measuring circuit, and a shunt resistance $R_s$, to take into account possible leakages, as those due to defect that could locally lower the Schottky barrier. Hence, eq. (1) becomes:

$$I = \frac{R_p}{R_s + R_p}\left\{I_0\left[\exp\left(\frac{q(V - R_s I)}{nkT}\right) - 1\right] - \frac{V}{R_p}\right\}. \quad (3)$$

Eq. (3) provides an excellent fit to the forward current, but underestimates the reverse saturation current, which results about an order of magnitude higher than expected (Fig. 4 (a); the overall fit improves at lower temperatures (Figs 4 (b)).

The behavior of the junction under white LEDs illumination is shown in Fig. 4 (d): The forward characteristic remains unaffected (apart the appearance of an open circuit voltage of ~0.2 eV due to photovoltaic effect), while the reverse current dramatically increases reaching values higher than the forward current. This unexpected behavior, together with the anomalous dark leakage, indicates that other components add to the reverse current of the Gr/Si junction.

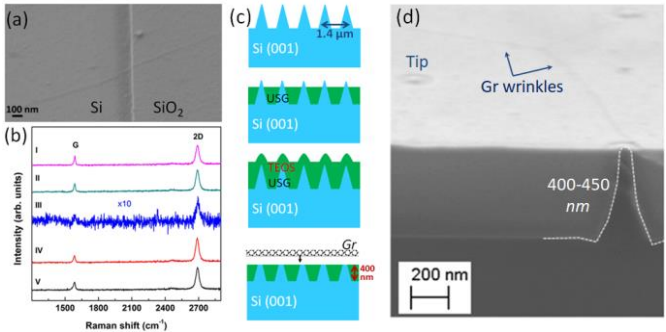

Figure 3. (a) SEM image of the flat Gr/Si junction showing graphene extending from the exposed to the $SiO_2$ covered Si region. (b) Micro-Raman spectra in different locations of the $SiO_2$ area (I, II, IV, V) and on Si (III). The signal from region III is multiplied by 10 for clarity. (c) Fabrication process of the Gr/Si-tips junction. From top to bottom: RIE to etch the Si-tips, USG and BPSG dielectric deposition, CMP to uncover the tips, graphene transfer. (d) SEM image of the Si nanopatterned substrate covered by graphene. Tips (height: 400-450 nm, top diameter: 40-50 nm) are highlighted, together with typical graphene wrinkles caused by the transfer process.

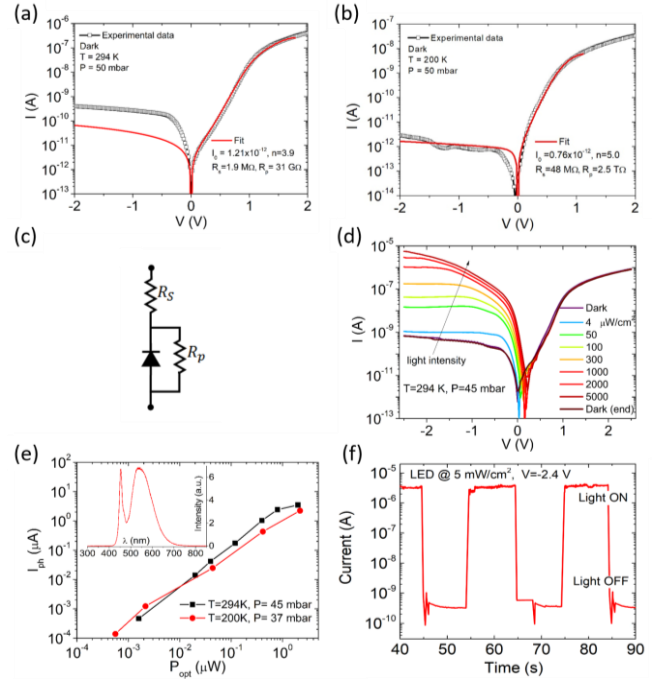

Figure 4. I-V characteristic at temperatures T=294 K (a) and T= 200 K (b) and fit according to a diode model (c) including a series ($R_s$) and shunt ($R_p$) resistance. (d) I-V characteristics for increasing white LEDs illumination intensity up to 5 mW/cm$^2$. (e) Photocurrent in reverse bias at -2.4 V and spectrum of the LED array (inset). (f) Transient behaviour under light on/off switching.

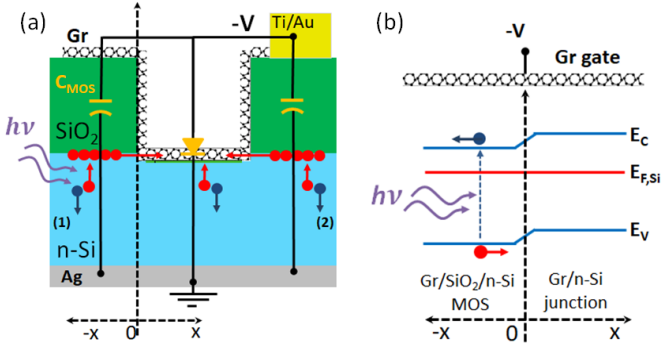

Figure 5. (a) Layout of the flat Gr/Si device showing two components: a MOS capacitor (Gr/SiO$_2$/Si) in parellel with a diode (Gr/Si junction). (b) Band diagram along the Si surface from the MOS to the diode region.

The responsivity of the device, defined as the ratio of the photocurrent ($I_{ph} = I_{light} - I_{dark}$) to the incident power, results $R = \frac{I_{ph}}{P_{opt}} = 2.5 \frac{A}{W}$ at ~3 mW/cm$^2$. This value exceeds the performance of solid-state photodetectors on the market or of similar Gr/Si devices [14]. Furthermore, Figs 4 (e) and 4 (f) show a photocurrent proportional to the incident power and a good transient behavior of the device.

To explain the experimental findings, we propose the model illustrated in Fig. 5 (a), in which both the part covered by SiO$_2$ and the junction region contribute to the observed optoelectronic behavior of the flat Gr/Si device. A large area Gr/SiO$_2$/Si MOS (Metal-Oxide-Semiconductor) capacitor surrounds the Gr/Si junction. In reverse bias, thermally or optically generated holes (minority carriers) are attracted to the SiO$_2$/Si side of the MOS capacitor. By diffusion, they reach the junction region and are swept by the electric field of the reverse biased junction, thus contributing to the reverse current. We point out that the band bending along the Si surface (Fig. 5 (b)), due to the gating effect of graphene [15], favors the drift of holes from the MOS to the junction region (where, as we will demonstrate, there is an inadvertent ultrathin dielectric layer).

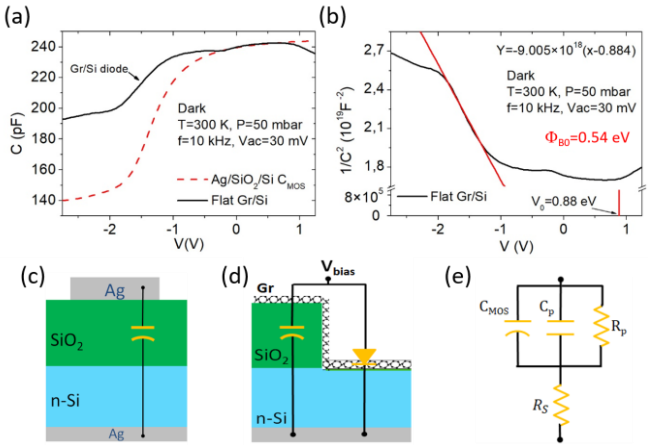

Figure 6. High frequency C-V measurement (a) and $\frac{1}{C^2}$ vs V plot (b) for the flat Gr/Si junction. An Ag/SiO$_2$/Si MOS test structure (c) is mesaured for comparison with the flat Gr/Si junction (d) which consists of the parellel of a MOS and a diode (e). In the C-V measurements, the reverse-biased diode is modelled as a capacitor, $C_p$, in parallel with a resistor $R_P$.

The model is supported by the high frequency (10 kHz) capacitance-voltage (C-V) measurements shown in Fig. 6. The C-V curve of an Ag/SiO$_2$/Si MOS test structure (Fig. 6 (c)) is compared with the C-V characteristic of flat Gr/Si device (Fig. 6 (d)), composed of the mentioned Gr/SiO$_2$/Si MOS and the Gr/Si diode (Fig. 6 (e)). As shown in Fig. 6 (a), the MOS test structure exhibits the typical accumulation region, with higher capacitance, in forward bias and the depletion region, with lower capacitance, in reverse bias. By contrast, for the Gr/Si device, the Gr/SiO$_2$/Si MOS component, which is dominant at positive voltage, is overcome by the capacitance of the depletion layer of the Gr/Si diode in reverse bias. This originates, for negative bias, the shoulder highlighted in the plot of Fig. 6 (a). The depletion layer capacitance of a diode, when an ultrathin interfacial oxide layer is considered, is expressed by [8,16]:

$$\frac{1}{C^2} = \frac{2\,n\,[n\,(\Phi_{b0} - \Phi_n - kT) - qV]}{A^2 q^2 \varepsilon_s N_d} \quad (4)$$

where n = 3.9 is extracted from the fit of Fig. 4 (a), $N_d = 4.5 \times 10^{14}$ cm$^{-3}$ is the doping density of Si, $\varepsilon_s = 11.7\varepsilon_0$ is the Si permittivity, and $\Phi_n = kT\ln\frac{N_c}{N_d}$ with $N_c$ the effective density of states in the conduction band (= $2.9 \times 10^{19}$ cm$^{-3}$ at T = 300K). The x-intercept, $V_0$, of the straight-line fitting the $1/C^2 - V$ curve, shown in Fig. 6 (b), is used to evaluate the barrier height as

$$\Phi_{b0} = \frac{V_0}{n} + kT\ln\left(\frac{N_c}{N_d}\right) + kT \approx 0.54 \text{ eV} \quad (5)$$

a value in good agreement with the barrier height of 0.52 eV extracted with the Richardson method, based on I-V measurements shown in Fig. 7. The forward part of the I-V curves at different temperatures is used to extract the reverse current at zero bias, $I_0(T)$, which, according to eq. (2), is plotted as $\ln\frac{I_0}{T^2}$ vs $\frac{1}{T}$ to extract the barrier height [1,9,10]. The y-intercept of the Richardson plot, which is shown in Fig. 7 (b), yields a Richardson constant of $3.9 \times 10^{-5}$ Acm$^{-2}$K$^{-2}$ that is several orders of magnitude lower than the theoretical value. We attribute the discrepancy to the inadvertent presence of a native oxide layer at Gr/Si interface. Such insulating layer modifies eq. (2) by a tunneling factor [8]:

$$I_0 = AA^*T^2 \exp(-\chi^{1/2}\delta)\exp\left(-\frac{\Phi_{b0}}{kT}\right) \quad (6)$$

where $\chi \approx 3$eV is the mean barrier height and $\delta$[Å] is the thickness of the layer.

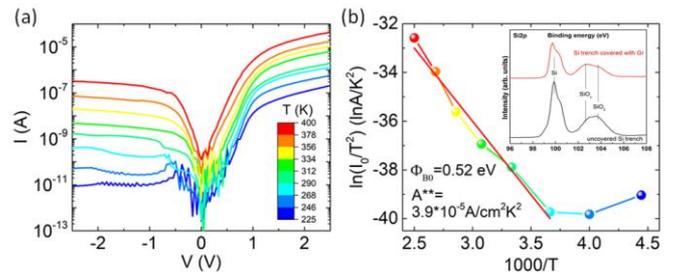

Figure 7. (a) I-V curves at different temperature and (b) Richardson plot of the flat Gr/Si device. The inset shows XPS spectra which confirm the presence of a sub-oxide layer (SiO$_x$, 1<x<2) below graphene at the Gr/Si interface.

Hence, a modified effective Richardson constant, $A^{**} = A^* \exp(-\chi^{1/2}\delta)$, where $A^{**}$ is the measured value and $A^*$ the theoretical one, can be used to estimate the oxide thickness as ~8 Å. Indeed, the presence of a sub-oxide $SiO_x$ layer below graphene with thickness ~1 nm is confirmed by the X-ray photoelectron spectra (inset of Fig. 7 (b)).

## V. GR/SI-TIPS JUNCTION: RESULTS AND DISCUSSION

We fabricated the device with graphene on patterned Si to further improve the photoresponse. The nanotips patterning has three major advantages: 1. The textured substrate favors light absorption through multiple reflections; 2. The reduced area favors barrier uniformity by reducing the probability of including defects; 3. More importantly, the field amplification at the top of the tips (Fig. 8 (a)), that is at the Gr/Si junction, enhances photocharge separation and can enable internal gain by avalanche multiplication.

A remarkable feature of the Gr/Si-tips junction is the linear dependence of the Schottky barrier height on the applied voltage, extracted from the I-V characteristics at different temperatures, as shown in Fig. 8 (c) [10]. The barrier increases (decreases) in forward (reverse) bias, related to the properties of graphene as explained previously, and expressed as

$$\Phi_B(V) = \Phi_{B0} + \gamma q(V - R_s I), \qquad (7)$$

where $\gamma$ is the slope of the fitting straight line.

We notice that the slightly lower value of the Schottky barrier at zero bias (0.36 eV vs 0.52 eV) with respect to the previous device can be attributed to the image force barrier lowering, which is more pronounced here, being proportional to the fourth root of substrate doping concentration ($N_d \sim 10^{18}$ cm$^{-3}$ for the Gr/Si-tips device).

The bias-modulation of the barrier can be considered in the diode equation by redefining the ideality factor as [10,16]:

$$\frac{1}{m} = \frac{1}{n} - \frac{\partial \Phi_B}{q \, \partial(V - R_s I)} = \frac{1}{n} - \gamma. \qquad (8)$$

Including eq. (8) in eq. (1) leads to the expression:

$$I = I_0 \left[ e^{q(V-R_s I)/mkT} - 1 \right] - I_0 \left[ e^{-q(V-R_s I)/\gamma^{-1}kT} - 1 \right] \quad (9)$$

that corresponds to the parallel of two opposite diodes, with $m$ and $\gamma^{-1}$ ideal factors respectively, in series with the resistance $R_s$, as shown in the inset of Fig. 8 (c)). Eq. (9) provides an excellent fit to the experimental data, as shown in Fig. 8 (c). From a physical viewpoint, the eq. (9) model considers both the injection of electrons from graphene to Si due to the lowering Schottky barrier in reverse barrier (Fig. 8 (d)) and the injection of electrons from Si to graphene due to the lowering of the Si depletion layer barrier in forward bias (Fig. 8 (e)).

Since the Gr/Si-tips device consists of more than 3 millions tips (corresponding to a total junction area of $6.079 \times 10^{-5}$ cm$^2$), the homogeneity of the Schottky barrier is an important parameter to check [17]. Assuming that the spatial distribution of the barrier follows a Gaussian distribution, with average value $\Phi_{Bm}$ and standard deviation $\sigma_B$, $P(\Phi_B) = \frac{1}{\sqrt{2\pi}\sigma_B} \exp\left(-\frac{(\Phi_B - \Phi_{Bm})^2}{2\sigma_B^2}\right)$, as shown in Fig. 9 (a), then the dependence of the barrier height on temperature is [18]:

$$\Phi_B = \Phi_{Bm} - \frac{q\sigma_B^2}{2kT}. \qquad (10)$$

The experimental data, reported in Fig. 9 (b), follow this model and yield a $\sigma_B = 74$ meV lower than typical values obtained for flat Gr/Si junctions and comparable to the barrier inhomogeneity of industrial M/S Schottky diodes.

The photoresponse of the device was measured under the same conditions of the flat Gr/Si device. Figs 9 (c) and 9 (d) show the response to the white LED light: Taking into account the effective junction area, this photocurrent corresponds to a responsivity $\geq 3$ A/W at 3 mW/cm$^2$. For this device the graphene gating effect is negligible, since the $SiO_2$ layer is too thick (400-450 nm). The record responsivity is due to the field amplification on the top of the nanotips, i.e. in the junction area, which facilitates photocharge separation and likely causes internal gain through impact ionization [10].

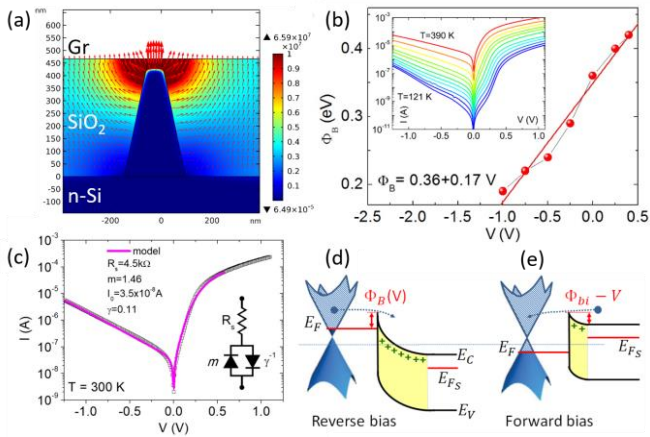

Figure 8. (a) Simulated electric field in the space around a Si-tip at V=-1V; the field arrows are in logarithmic scale (simulation by COMSOL software). (b) Schottky barrier height as a function of the applied bias; the inset shows the measured I-V characteristics at temperatures stepping from 120 K to 390 K. (c) Fitting of a two-diode model to the I-V curve at room temperature and equivalent circuit (inset). Electron injection due to the modulation (d) of the Schottky barrier height in reverse bias and (e) of the depletion layer barrier in forward bias.

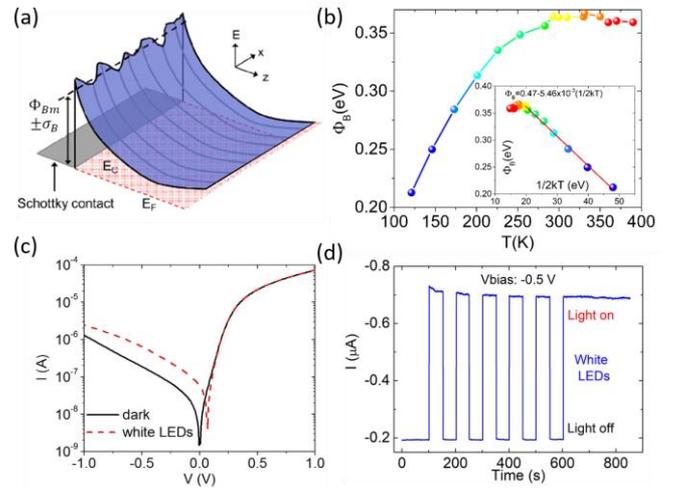

Figure 9. (a) Spatial inhomogeneity of a Schottky barrier. (b) Measured Schottky barrier height as function of the temperature. Photoresponse of the device to white LED light: (c) I-V characteristics and (d) transient response to on/off light switching.

## VI. Conclusion

We have reviewed key electrical features of the Gr/Si junction and focused on two specific devices. Although caused by different mechanisms, on both devices we have reported a responsivity exceeding that of present solid-state devices. This work unveils new physical phenomena occurring in Gr/Si photodiodes and demonstrates the high potential of such devices as next-generation photodetectors.

## Acknowledgment

We thank Materials Research Department, IHP-Microelectronics, Frankfurt Oder, Germany, for the fabrication of the devices and the general support to this project.

## References

[1] A. Di Bartolomeo, "Graphene Schottky diodes: An experimental review of the rectifying graphene/semiconductor heterojunction," *Physics Reports*, vol. 606, pp. 1-58, Jan. 2016

[2] S. Tongay, M. Lemaitre, X. Miao, B. Gila, B. R. Appleton, A. F. Hebard, "Rectification at Graphene-Semiconductor Interfaces: Zero-Gap Semiconductor-Based Diodes," *Physical Review X*, vol. 2, no. 1, art. no.. 011002, pp. 9, 2012.

[3] A. Di Bartolomeo, F. Giubileo, F. Romeo, P. Sabatino, G. Carapella, L. Iemmo, T. Schroeder, and G. Lupina, "Graphene field effect transistors with niobium contacts and asymmetric transfer characteristics," *Nanotechnology*, vol. 26, no. 47, art. no. 475202, pp. 9, Nov. 2015

[4] R. R. Nair, P. Blake, A. N. Grigorenko, K. S. Novoselov, T. J. Booth, T. Stauber, N. M. R. Peres, and A. K. Geim, "Fine Structure Constant Defines Visual Transparency of Graphene," *Science*, vol. 320, no. 5881, pp. 1308–1308, Jun. 2008.

[5] S. Riazimehr, D. Schneider, C. Yim, S. Kataria, V. Passi, A. Bablich, G. S. Duesberg and M. C. Lemme, "Spectral sensitivity of a graphene/silicon pn junction Photodetector," *EUROSOI-ULIS 2015 - 2015 Joint International EUROSOI Workshop and International Conference on Ultimate Integration on Silicon*, no. 7063777, pp. 77-80, Mar. 2015.

[6] I. Goykhman, U. Sassi, B. Desiatov, N. Mazurski, S. Milana, D. de Fazio, A. Eiden, J. Khurgin, J. Shappir, U. Levy, and A. C. Ferrari, "On-Chip Integrated, Silicon–Graphene Plasmonic Schottky Photodetector with High Responsivity and Avalanche Photogain," *Nano Letters*, vol. 16, no. 5, pp. 3005–3013, Apr. 2016.

[7] X. Wan, Y. Xu, H. Guo, K. Shehzad, A. Ali, Y. Liu, J. Yang, D. Dai, C.-T. Lin, L. Liu, H.-C. Cheng, F. Wang, X. Wang, H. Lu, W. Hu, X. Pi, Y. Dan, J. Luo, T. Hasan, X. Duan, X. Li, J. Xu, D. Yang, T. Ren and B. Yu, "A self-powered high-performance graphene/silicon ultraviolet photodetector with ultra-shallow junction: breaking the limit of silicon?," *npj 2D Materials and Applications*, vol. 1, art. no. 4, pp. 8, Apr. 2017.

[8] A. Di Bartolomeo, G. Luongo, F. Giubileo, N. Funicello, G. Niu, T. Schroeder, M. Lisker, G. Lupina, "Hybrid graphene/silicon Schottky photodiode with intrinsic gating effect", *2D Materials*, vol. 4, no. 2, art. no. 025075, pp. 11, Apr. 2017.

[9] G. Luongo, F. Giubileo, L. Genovese, L. Iemmo, N. Martucciello, A Di Bartolomeo "I-V and C-V Characterization of a High-Responsivity Graphene/Silicon Photodiode with Embedded MOS Capacitor," *Nanomaterials*, vol. 7, no. 7 art. no. 158, pp. 8, Jun. 2017.

[10] A. Di Bartolomeo, F. Giubileo, G. Luongo, L. Iemmo, N. Martucciello, G. Niu, M. Fraschke, O.Skibitzki, T. Schroeder, G. Lupina, "Tunable Schottky barrier and high responsivity in graphene/Si-nanotip optoelectronic device," *2D Materials*, vol. 4, no. 1, art. no. 015024, pp. 10, 2017.

[11] G. Lupina, J. Kitzmann, I. Costina, M. Lukosius, C. Wenger, A. Wolff, S. Vaziri, M. Östling, I. Pasternak, A. Krajewska, W. Strupinski, S. Kataria, A. Gahoi, M. C. Lemme, G. Ruhl, G. Zoth, O. Luxenhofer, W. Mehr, "Residual Metallic Contamination of Transferred Chemical Vapor Deposited Graphene," *ACS Nano*, vol. 9, no. 5, pp. 4776–85, Apr. 2015.

[12] O. Skibitzki, I. Prieto, R. Kozak, G. Capellini, P. Zaumseil, Y. A. R. Dasilva, M. D. Rossell, R. Erni, H. von Känel and T. Schroeder, "Structural and optical characterization of GaAs nano-crystals selectively grown on Si nano-tips by MOVPE," *Nanotechnology*, vol. 28, no. 13, art. no. 135301, pp. 10, Feb. 2017.

[13] F. Giubileo, A. Di Bartolomeo, "The role of contact resistance in graphene field-effect devices," *Progress in Surface Science*, vol. 92, no. 3, pp. 143-175, Aug. 2017.

[14] S. Riazimehr, S. Kataria, R. Bornemann, P. H. Bolívar, F. J. G. Ruiz, O. Engström, A. Godoy, and M. C. Lemme, "High Photocurrent in Gated Graphene−Silicon Hybrid Photodiodes," *ACS Photonics*, vol. 4, no. 6, pp. 1506–1514, Jun. 2017.

[15] A. Di Bartolomeo, F. Giubileo, L. Iemmo, F. Romeo, S. Russo, S. Unal, M. Passacantando, V. Grossi, and A. M. Cucolo, "Leakage and field emission in side-gate graphene field effect transistors," *Applied Physics Letters*, vol. 109, no. 2, art. no. 023510, pp. 1-5, Jul. 2016.

[16] M. S. Tyagi, "Physics of Schottky Barrier Junctions" in *Metal-Semiconductor Schottky Barrier Junctions and Their Applications*, B. L. Sharma ed., Boston, MA: Springer US, 1984, ch. 1, pp. 1–60.

[17] S.-J. Liang, W. Hu, A. Di Bartolomeo, S. Adam, and L.K. Ang, "A modified Schottky model for graphene-semiconductor (3D/2D) contact: A combined theoretical and experimental study," 62nd IEEE International Electron Devices Meeting, IEDM 2016; San Francisco; United States; 3-7 December 2016; *Technical Digest - International Electron Devices Meeting, IEDM*, vol. 2016, art. no. 7838416, pp. 14.4.1-14.4.4, Jan. 2017.

[18] J. H. Werner and H. H. Guttler, "Barrier Inhomogeneities at Schottky Contacts," *Journal of Applied Physics*, vol. 69, no. 3, pp. 1522-1533, 1991.